\documentclass[a4paper]{aa}
\usepackage{graphicx}
\usepackage{txfonts}
\begin{document}
\newcommand{\pp}{$\psi$\,Phe}
\newcommand{\rsun}{$R_{\odot}$}
\newcommand{\msun}{$M_{\odot}$}
\newcommand{\lsun}{$L_{\odot}$}
\newcommand{\teff}{$T_{\rm eff}$}
\newcommand{\adross}{$\theta_{\rm Ross}$}
\newcommand{\adld}{$\theta_{\rm LD}$}
\title{Tests of stellar model atmospheres by optical interferometry}
\subtitle{VLTI/VINCI limb-darkening measurements of the 
M4 giant \pp\thanks{Based on public data released 
from the European Southern Observatory VLTI
obtained from the ESO/ST-ECF 
Science Archive Facility. The VLTI was operated with the commissioning 
instrument VINCI and the MONA beam combiner.}}
\titlerunning{VLTI/VINCI limb-darkening measurements of \pp}
\author{
M. Wittkowski \inst{1} \and
J. P. Aufdenberg \inst{2} \and
P. Kervella \inst{3} 
}
\offprints{M.~Wittkowski, \email{mwittkow@eso.org}}
\institute{European Southern Observatory, Karl-Schwarzschild-Str. 2,
85748 Garching bei M\"unchen, Germany\\
\email{mwittkow@eso.org}
\and
Harvard-Smithsonian Center for Astrophysics, 60 Garden Street, 
Mail Stop 15, Cambridge, MA 02138, USA\\
\email{jaufdenberg@cfa.harvard.edu}
\and European Southern Observatory, Casilla 19001, Santiago 19, Chile;
\email{pkervell@eso.org}
}
\date{Received \dots; accepted \dots}
\abstract{
We present $K$-band interferometric measurements of the 
limb-darkened (LD) intensity profile of the M\,4 giant star $\psi$\,Phoenicis 
obtained with the Very Large Telescope Interferometer (VLTI) and its 
commissioning instrument VINCI.  High-precision squared visibility 
amplitudes in the second lobe of the 
visibility function were obtained employing two 
8.2\,m Unit Telescopes (UTs). This succeeded one month after 
light from UTs was first combined for interferometric fringes.
In addition, we sampled the visibility function at small spatial 
frequencies using the 40\,cm test siderostats. Our measurement
constrains the diameter of the star as well as its center-to-limb 
intensity variation (CLV).
We construct a spherical hydrostatic {\tt PHOENIX} model atmosphere based on 
spectrophotometric data from the literature and confront its CLV prediction 
with our interferometric measurement. 
We compare as well CLV predictions by plane-parallel 
hydrostatic {\tt PHOENIX}, {\tt ATLAS\,9}, and {\tt ATLAS\,12} models.
We find that the Rosseland angular diameter as predicted
by comparison of the spherical {\tt PHOENIX} model with spectrophotometry
is in good agreement with our interferometric diameter measurement.
The shape of our measured visibility function in the second lobe
is consistent with all considered {\tt PHOENIX} and {\tt ATLAS} 
model predictions, and significantly different from uniform disk (UD)
and fully darkened disk (FDD) models.
We derive high-precision fundamental parameters for \pp, 
namely a Rosseland angular diameter of 8.13 $\pm$ 0.2 mas, with the
Hipparcos parallax corresponding 
to a Rosseland linear radius $R$ of 86 $\pm$ 3 R$_\odot$, 
and an effective temperature of 3550 $\pm$ 50 K, with $R$ corresponding
to a luminosity of $\log L$/\lsun=3.02 $\pm$ 0.06. Together with 
evolutionary models, these values are consistent with a mass
of 1.3 $\pm$ 0.2 \msun, and a surface gravity of 
$\log g =$ 0.68 $\pm$ 0.11.
\keywords{
Techniques: interferometric --
Stars: atmospheres --
Stars: fundamental parameters --
Stars: late-type --
Stars: individual: \object{$\psi$\,Phe}
}
}
\maketitle
\section{Introduction}
Stellar atmosphere models predict the spectrum emerging from every
point of a stellar disk. However, model atmospheres are usually only 
constrained by comparison to integrated stellar spectra. 
Optical interferometry has proven its capability to go beyond this 
principal test of the predicted flux, and to probe
the wavelength-dependent center-to-limb intensity variation (CLV) 
across the stellar disk. 
In addition, the measurement of the stellar angular diameter together 
with the bolometric flux is the primary measure of the effective 
temperature, one of the most important parameters for modeling 
stellar atmospheres and stellar evolution. 

Further tests of stellar atmosphere models by interferometric
observations help to improve the reliability of results in all areas
of astrophysics where such models are used.
Cool giants and AGB stars are of interest for atmosphere modeling since they 
allow the study of extended stellar atmospheres and the
stellar mass-loss process. 
Red giants are also used as probes of the chemical enrichment history of 
nearby galaxies through detailed abundance measurements of the 
Calcium infrared triplet which
rely on model atmospheres.

However, the required direct measurements of stellar intensity profiles
are among the most challenging programs in modern optical interferometry.
Since more than one resolution element across the stellar disk is needed
to determine surface structure parameters beyond diameters, 
the long baselines needed to
obtain this resolution also produce very low visibility amplitudes
corresponding to vanishing fringe contrasts.
Consequently, direct interferometric limb-darkening observations of stars
with compact atmospheres, i.e. visibility measurements in the 2nd lobe, 
have so far been limited to a small number of stars 
(including Hanbury Brown et al. \cite{hanbury}; 
Di Benedetto \& Foy \cite{benedetto};
Quirrenbach et al. \cite{quirrenbach};
Burns et al. \cite{burns};
Hajian et al. \cite{hajian};
Wittkowski et al. \cite{wittkowski1}).
For stars with extended atmospheres, described by for instance 
Gaussian-type or two-component-type CLVs, measurements of high to medium 
spatial frequencies of the 1st lobe of the visibility function
may lead to CLV constraints as well 
(see, e.g. Haniff et al. \cite{haniff}; Perrin et al. \cite{perrin}). Furthermore, lunar occultation
measurements may also enable a reconstruction of the CLV (e.g. Richichi
et al. \cite{richichi}).
Recent optical multi-wavelength measurements 
of the cool giants $\gamma$\,Sge and BY\,Boo 
(Wittkowski et al. \cite{wittkowski1}) succeeded not only in 
directly detecting the limb-darkening
effect, but also in constraining {\tt ATLAS\,9} (Kurucz \cite{kurucz})
model atmosphere parameters. Aufdenberg \& Hauschildt (\cite{aufdenberg2})
showed that these $\gamma$\,Sge interferometric measurements
and $\gamma$\,Sge spectroscopic measurements both compare well with
predictions by the same spherical
{\tt PHOENIX} model atmosphere (Hauschildt et al. \cite{hauschildt1}).

Here, we present limb-darkening observations of the M\,4 giant star
\pp\ (\object{HD\,11695}, HR\,555, FK5\,67, HIP\,8837), 
obtained during the commissioning period of the ESO Very Large Telescope
Interferometer (VLTI) with its commissioning instrument VINCI. 
VINCI is operated with one near-infrared $K$-band filter and could not 
provide measurements at different spectral bands.
Spectrally resolved VLTI measurements are planned
with the upcoming instruments AMBER (Petrov et al. \cite{amber})
for near-infrared wavelengths, and MIDI (Leinert et al. \cite{midi})
for mid-infrared wavelengths.
We construct a spherical hydrostatic {\tt PHOENIX} model atmosphere based 
on spectrophotometric data from the literature. This is the usual procedure 
to obtain a model atmosphere since additional interferometric observations 
are usually not available. 
This resulting atmosphere model predicts the LD 
intensity profile, a prediction that we confront with our 
interferometric measurement in order to test it. 
Agreement of the model prediction and our measurement 
increases confidence in atmosphere modeling for cool giants.
In addition, we compare predictions of a plane-parallel
{\tt PHOENIX} model atmosphere as well as 
of standard plane-parallel {\tt ATLAS\,9} (Kurucz \cite{kurucz}) 
and {\tt ATLAS\,12} (Kurucz \cite{kurucz12}, \cite{kurucznew}) 
atmosphere models.

The observations and methods presented here are also a precursor 
for more detailed limb-darkening measurements and
observations of other stellar surface features.
The feasibility to derive stellar surface structure parameters beyond 
diameters and limb-darkening using VLTI and its scientific instruments has
been studied by, e.g., von der L\"uhe (\cite{vonderluehe}), 
Jankov et al. (\cite{jankov}), and
Wittkowski et al. (\cite{wittkowski2}).
\section{Characteristics of \pp}
\subsection{\pp\ -- A spectroscopic binary ?}
% RV   eRV   Ref
% +8.9  ?     Lunt 1919 Lick  ``obvious of supspected variable''
% +3.8  4.2   Lunt 1919 Cape  ``obvious of supspected variable''
% +1.0  30%   Wilson 1953 (+6.8 (4 from Lick) & +0.0 (15 from Cape))
% +5.0  0.4   Evans Menzies Stoy 1957
% +1.0  c     Eggen 1960 (clearly from Wilson 1953)
% +1.5        Evans 1967 IAU Symp. 30, 57 (ed. A.H. BATTEN & J.F. HEARD, Academic Press
%  0.0        Eggen 1973
% +5.6        Jones 1972
% +3.0  c     Crampton and Evans 1973 
% +1.0  var   Augensen and Buscombe 1978 (clearly from Wilson)
% +6.0  0.9   Jones and Fisher 1984
The Hipparcos Catalogue (Perryman \& ESA \cite{esa}) notes
in the spectral type column
that \pp\ is a spectroscopic binary (``M4III SB''). \pp\ is hardly worthy 
of the spectroscopic binary
designation.  There is no mention of this star 
in the ``Catalogue of the orbital elements of spectroscopic binary systems''
(Batten et al. \cite{batten}).
Likely sources for the ``SB'' note are Houk (\cite{houk}) and 
Wilson (\cite{wilson}).
The radial velocity from Wilson (\cite{wilson}) is noted with a (poor) 
grade ``c'' and the designation ``SB''.
The ``SB'' appears to be based on a note by Lunt (\cite{lunt}): ``either
obviously variable in velocity or suspected''.  
Since then, a small number of radial velocity measurements
(Evans et al. \cite{evans}; Jones \cite{jones}; 
Crampton \& Evans \cite{crampton}; Jones \& Fischer \cite{jones2}) 
have failed to show variations in \pp's radial velocity.

While the variability of \pp's radial velocity has not been
convincingly established, it is clear that \pp\ is a
small-amplitude ($\Delta V = \pm 0.1$), short-period ($\simeq$ 30 d),
photometric variable star (Eggen \cite{eggen}). Our high spatial frequency
observations (see Sect.~\ref{sec:observations}) cover two days 
or $\sim$ 1/15 of the photometric period. 
We are not able to establish the photometric phase of the
interferometric observations since there was no
follow-up to Eggen's original photometry.
\subsection{The bolometric flux of \pp}
\label{sec:fbol}
We determine the bolometric flux of \pp\ to be 
$F_{\rm Bol} = (3.2 \pm 0.3)\times
10^{-9}$ W\,m$^{-2}$, derived from a spline fit and integration of the
available spectrophotometry.
We have used the narrow-band 
spectrophotometric data ($\Delta\lambda$ = 2.5 nm, 400 nm -- 700 nm)
from Burnashev (\cite{burnashev}) and the medium- and broad-band 
optical/near-IR photometric data from Johnson \& Mitchell 
(\cite{johnson2}, \cite{johnson3}) and Feast et al. (\cite{feast}).  
The absolute calibration of the 13-color photometry is taken 
from Johnson \& Mitchell (\cite{johnson2}, \cite{johnson3}).
The absolute calibration of JKLM photometry is from 
Johnson (\cite{johnson1}) while
the H-band calibration is from Bessel \& Brett
(\cite{bessel}). We have assumed a uniform 5\% error for all absolute
fluxes.  The color excess towards \pp\ is very low due its high
galactic latitude (b = --67$^\circ$).  The {\it COBE} dust maps
(Schlegel et al. \cite{schlegel}) indicate that 
the {\it maximum} $E(B-V)$ value for this line-of-sight is 0.026 
and we take this as the upper limit on
$E(B-V)$. This uncertainty in the interstellar extinction is
included in the derived bolometric flux.  The $V$-band variability of
\pp, which we have not included, may be a source of additional
uncertainty in $F_{\rm bol}$.
\begin{table*}
\centering
\caption{Properties of \pp. The upper table lists the adopted literature 
values (visual magnitude, near-infrared $K$-band magnitude, spectral type,
metallicity, parallax, bolometric flux).
The lower table compares the Rosseland angular diameter
and the effective temperature with their $1\,\sigma$ errors as derived in this 
work using different methods, as well as derived parameters 
(effective temperature, 
Rosseland radius, luminosity, mass, surface gravity) with their
errors.}
\label{tab:psipheprop}
\begin{tabular}{lllllll}
    & $V$ & $K$ & Sp. Type & $[Z/Z_\odot]$ & $\pi$ & $F_{\rm bol}$ \\\hline
\pp & 4.3-4.5 (period $\simeq$ 30\,d) $^1$ & -0.63 $\pm$ 0.02 $^2$ & M4 III $^3$ &  0.25 $\pm$ 0.1 $^4$ & 10.15 $\pm$ 0.15 mas $^5$ & 
$(3.2 \pm 0.3)\times 10^{-9}$ W\,m$^{-2}$ $^6$ \\
\end{tabular}\\

$^1$ Eggen (\cite{eggen});
$^2$ Gezari et al. (\cite{gezari});
$^3$ Houk (\cite{houk});
$^4$ Feast et al. (\cite{feast}) ;
$^5$ Perryman \& ESA (\cite{esa});
$^6$ see Sect.\,\ref{sec:fbol} \\[1.5ex]

\begin{tabular}{lccccccc}
& Photometric & Fit of & \multicolumn{4}{c}{Fit of}   &Final  \\
& estimates   & spherical                 & spherical                     & plane-parallel     & plane-parallel             & plane-parallel              & values \\
&   & {\tt PHOENIX} model       & {\tt PHOENIX}                 & {\tt PHOENIX}      &{\tt ATLAS\,9}              & {\tt ATLAS\,12}             & \\
&             & to spectrophotometry      & \multicolumn{4}{c}{model to our interferometric data}  &   \\
& Sect.~\ref{sec:phot} & Sect.~\ref{sec:phoenixspec} & Sect.~\ref{sec:comp} & Sect.~\ref{sec:comp} & Sect.~\ref{sec:comp} & Sect.~\ref{sec:comp} & Sect.~\ref{sec:discussion} \\
\hline
\adross\ (mas) & 8.0 $\pm$ 0.8 $^a$        &  8.0 $\pm$ 0.4                & 8.13 $\pm$ 0.2 & 8.17 $\pm$ 0.2 & 8.24 $\pm$ 0.2 & 8.19 $\pm$ 0.2 & 8.13 $\pm$ 0.2$^g$           \\[0.5ex]
\teff\ (K)   &                           &  3550 $\pm$ 50                &                &         &       &                & 3550 $\pm$ 50 $^h$           \\[2ex]
\teff\ (K)   & 3500 $\pm$ 260 $^b$       &  3500 $\pm$ 170 $^b$          &                &         &       &                & 3472 $\pm$ 125 $^b$          \\[0.5ex]
$R$/\rsun     & 85 $\pm$ 10 $^c$          &  85 $\pm$ 6 $^c$          &                &         &       &                & 86 $\pm$ 3 $^c$              \\[0.5ex]
$\log L$/\lsun     & 2.99 $\pm$ 0.23 $^d$       &  3.01 $\pm$ 0.08 $^d$         &                &         &       &                & 3.02 $\pm$ 0.06 $^d$         \\[1.0ex]
$M$/\msun     & 1.2 $^{+0.8}_{-0.6}$ $^e$ &  1.2 $\pm$ 0.4 $^e$           &                &         &       &                & 1.3 $\pm$ 0.2 $^e$           \\[0.5ex]
$\log g$ (cgs)  & 0.66 $^{+0.33}_{-0.57}$ $^f$ &  0.66 $^{+0.18}_{-0.23}$ $^f$ &                &         &       &                & 0.68 $^{+0.10}_{-0.11}$ $^f$ \\
\end{tabular}\\[1ex]

$^a$ Using the calibration by Dyck et al. (\cite{dyck}) based on the 
spectral type and the $K$-magnitude;
$^b$ With \adross\ and $F_{\rm bol}$; 
$^c$ with \adross\ and $\pi$; 
$^d$ with $R$ and \teff;
$^e$ with $L$, \teff, and the evolutionary tracks by Girardi et al. (\cite{girardi}), see Fig. \ref{fig:hr};
$^f$ with $M$ and $R$, $^g$ from spherical {\tt PHOENIX} model fit to 
interferometric data,
$^h$ from spherical {\tt PHOENIX} model fit to spectrophotometry.
\end{table*}
\subsection{A priori estimate of fundamental parameters}
\label{sec:phot}
In order to decide on the grid of \pp's mass, radius, and effective 
temperature to explore, we roughly estimate these parameters.
The diameter of \pp\ has never been measured before by a direct technique. 
Using the calibration of Dyck et al. (\cite{dyck}), based on the 
spectral type and the $K$-magnitude, we derive 
a Rosseland angular diameter of 8.0 $\pm$ 0.8 mas. 
Here, the Rosseland radius is defined as the radius at which the 
Rosseland optical depth equals unity, a definition that we follow
in this article.

With our value for $F_{\rm bol}$, \teff\ is 
constrained to 3500 $\pm$ 260 K. With the Hipparcos parallax, 
the linear Rosseland radius is derived to 85 $\pm$ 10 \rsun, 
and the luminosity to $\log L$/\lsun = 2.99 $\pm$ 0.23. With these values we
can place \pp\ on the theoretical H-R diagram (see Fig.\ref{fig:hr}). 
Together with the evolutionary tracks of Girardi et al. (\cite{girardi}),
these values for \teff\ and $L$ give a 
mass estimate of 1.2 $^{+0.8}_{-0.6}$ \msun, and a $\log g$ estimate
of 0.66 $^{+0.33}_{-0.57}$.
Feast et al. (\cite{feast}) derive a metallicity 
$[Z/Z_\odot]$ of 0.25 $\pm$ 0.08, based on JHK photometry and atmosphere
models by Bessell et al. (\cite{bessell}). However, they argue that 
this offset might be an artifact since $[Z/Z_\odot]=0$ should be
appropriate for local M giants. The values given in this section
are summarized in Table~\ref{tab:psipheprop}.
\section{Calculation of atmosphere models for \pp}
\label{sec:models}
The modeling of atmospheres of cool giant stars, such as \pp, 
is complicated by two effects, the treatment of molecular
opacities (molecules form due to the low temperatures), and the
effects of their spherical extension (cf. Plez et al. \cite{plez}; 
Hauschildt et al. \cite{hauschildt2}). The appropriateness  of using
a plane-parallel model is based on the photospheric scale height
relative to the radius of the star. For the sun, with a photospheric 
thickness of $\sim$\,1000\,km, the extension of the photosphere relative 
to the solar radius is $\sim$\,0.1\%. For \pp\ the predicted photospheric
thickness is $\sim$\,4,000,000\,km and with a radius of $\sim$\,85\,\rsun, 
the extension is $\sim$\,6\%. Atmosphere models for M giants including 
spherical extension effects have been discussed by, e.g.,
Scholz \& Tsuji (\cite{scholztsuji}); Scholz (\cite{scholz});
Bessell et al. (\cite{bessell},\cite{bessell91});
Plez et al. (\cite{plez}); Hofmann \& Scholz (\cite{hofmann}); and 
Hauschildt et al. (\cite{hauschildt2}).
Here, we employ hydrostatic spherically symmetric {\tt PHOENIX} models
including state-of-the-art treatment of molecular opacities. These models
are based on the {\tt PHOENIX/NextGen} models 
by Hauschildt et al. (\cite{hauschildt2}).
We compare our observational results as well to a hydrostatic plane-parallel
{\tt PHOENIX} model.
Tabulated model CLVs based on hydrostatic 
plane-parallel ATLAS\,9 models are publicly 
available (Kurucz \cite{kurucz}) and often used for the 
interpretation of interferometric measurements of different stars 
including cool giants. These models are also used to parametrize 
LD profiles 
(for instance, Claret \cite{claret1}, \cite{claret2}; 
Davis et al. \cite{davis}). We compare the predictions by these 
standard models as well. Finally, we use a hydrostatic 
plane-parallel ATLAS\,12 model (Kurucz \cite{kurucz12}, \cite{kurucznew})
which, compared to the ATLAS\,9 models, includes an improved treatment 
of the molecular opacities, but no spherical extension effects.
\subsection{{\tt PHOENIX} model atmospheres}
\label{sec:phoenix}
We have computed new, fully line-blanketed (7$\times 10^5$ atomic
lines and 9$\times 10^7$ molecular lines), spherical, hydrostatic
atmosphere models with solar photospheric abundances 
(Grevesse \& Noels \cite{grevesse})
with version 13 of the PHOENIX code (for a general description see
Hauschildt \& Baron \cite{hauschildt1}).  
The three most important 
input parameters for our spherical models are the effective temperature, 
the surface gravity, and the stellar mass. Our new models start from 
the {\tt NextGen} model grid of 
Hauschildt et al. (\cite{hauschildt2}) with a stellar 
mass of 0.5\,$M_\odot$. 
The microtubulence for all our new models is 2\,km\,s$^{-1}$,  
as adopted for the NextGen grid.
The new model structures are iteratively converged with
temperature corrections following energy conservation for convective
and radiative equilibrium.  We have computed a grid of 63 new models 
at seven temperatures: 3450, 3475, 3500, 3525, 3550, 3575, 3600 K; three
gravities: 0.5, 0.7, 1.0; and three masses 0.7, 1.0, 1.3 M$_\odot$.
In addition we have computed 27 models at 
masses 2.0, 2.5, 3.0 M$_\odot$ for temperatures 3500, 3550, 3600 K
and gravities 0.5, 0.7. 1.0 in order to study the effect of higher 
masses; as well as 7 additional models at gravities 0.0, 0.25, 
0.5, 0.7, 1.0, 1.25, 1.5 for mass 1.0\,M$_\odot$ and temperature 3550\,K 
in order to study the effects of different gravities over a larger range.
For each model, we tabulate the intensity profile at 64 angles for
wavelengths from 1.8\,$\mu$m to 2.5\,$\mu$m in steps of 0.5\,nm.
The mass and input surface gravity fix the reference radius (at
$\tau$(Rosseland)=1) at which the effective temperature is defined.
Furthermore, the mass effectively regulates the relative extension of
the atmosphere for fixed values of the effective temperature and
gravity.  In hydrostatic equilibrium, the thickness of the atmosphere
relative to the overall radius increases as the mass decreases.  One
way to quantify the relative extension or degree of compactness of a
hydrostatic stellar atmosphere is to compare the gas-pressure scale
height with the stellar radius (see Bessell et al. \cite{bessell91};
Baschek et al. \cite{baschek}).  In extended
atmospheres the pressure scale height is a significant fraction of the
stellar radius, thus the ratio of these quantities is a useful
quantity.  For a fixed gravity, the stellar mass does not directly
affect the pressure scale height, $H_P = \frac{\mathcal{R}T}{\mu
g_{\rm eff}}$, where $\mathcal{R}$ is the gas constant, $T$ is the
temperature, $\mu$ is the mean molecular weight, and $g_{\rm eff}$ is
the effective gravity which includes radiative and turbulent
acceleration terms.  On the other hand, the stellar mass does directly
affect the stellar radius, $R = \sqrt{GM/g}$, where $G$ is the
gravitational constant, $M$ is the stellar mass, and $g$ is the
gravity. Hence the ratio $H_P/R$ decreases with increasing mass for a
fixed value of gravity.  Therefore, in the hydrostatic context,
spherical atmospheres for massive stars are more compact than for
lower mass stars with the same effective temperature and gravity.
As a result the mass can affect synthetic limb intensity profiles and 
therefore can be an important parameter when comparing spherical model
atmosphere predictions to interferometric data.
Angular diameter, temperature, gravity, and mass are, of course, 
dependent parameters and can not be independently explored over 
large ranges. 

For comparison, we use as well
a hydrostatic plane-parallel {\tt PHOENIX} model with an effective
temperature of 3550\,K and $\log g = 0.7$. The intensity profile is
tabulated at 64 angles for wavelengths from 1.8\,$\mu$m to 2.5\,$\mu$m 
in steps of 0.5\,nm, as for the spherical models.
\begin{figure}
\centering
\resizebox{1\hsize}{!}{\includegraphics[angle=0]{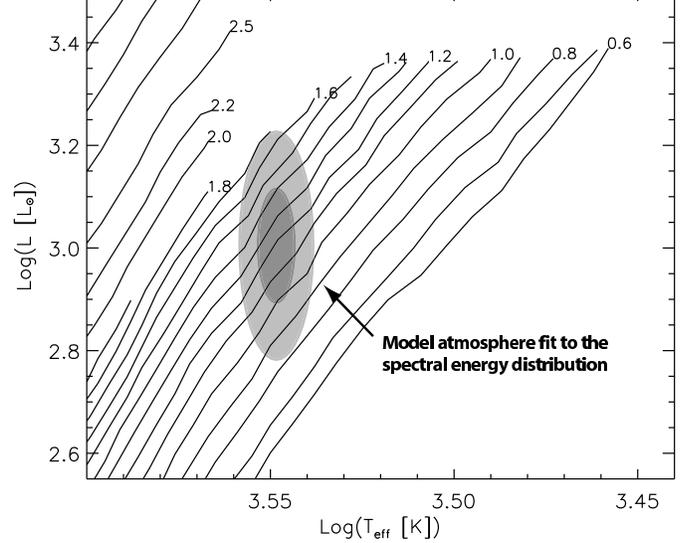}}
\caption{The predicted regions for \pp\ in the Hertzsprung-Russell 
diagram from the model atmosphere fitting to the observed spectral 
energy distribution.  
The method provides values for \teff\ and
\adross, and together with the parallax, a luminosity.
The error ellipses 
show 1- and 2-sigma zones.  Also shown are the evolutionary tracks of
Girardi et al. (\cite{girardi}) for masses $M > 0.6$\,\msun.}
\label{fig:hr}
\end{figure}
\begin{figure*}
\centering
\resizebox{1\hsize}{!}{\includegraphics[angle=90]{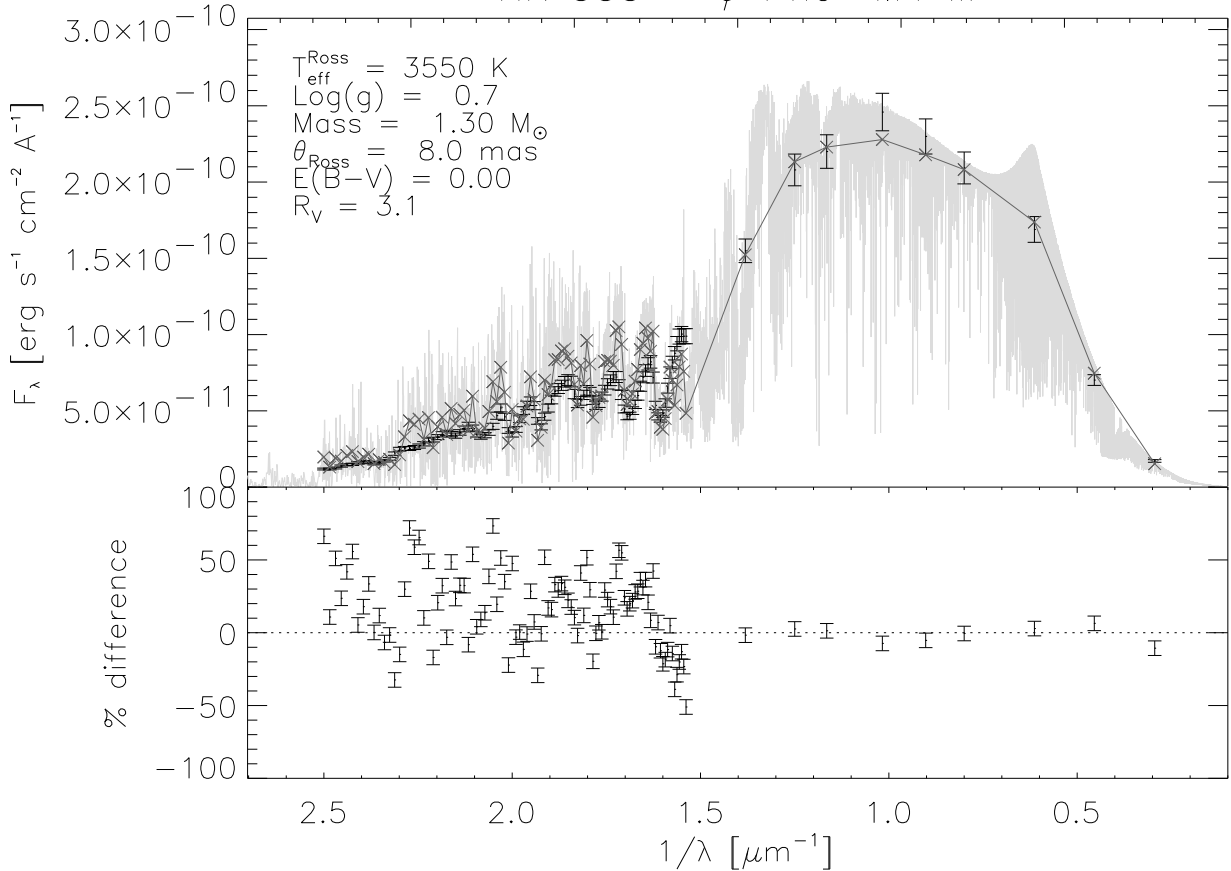}}
\caption{(Top) Spectral energy distribution of \pp\ (error bars)
compared with synthetic photometry ('x's) for the narrow and
broad-band photometry derived from a high-resolution model spectrum
(spherical {\tt PHOENIX} model) shown in gray. (Bottom) The percentage 
difference between the observed and synthetic photometry in each 
wavelength bin.
The model parameters are derived by a least-squares fit 
to only the 13-color photometry (Johnson \& Mitchell \cite{johnson2}),
while this Fig. shows the
model comparision to all available data, as described in the text.
}
\label{fig:sed}
\end{figure*}
\subsection{Comparison of {\tt PHOENIX} models with spectrophotometry}
\label{sec:phoenixspec}
The effective temperature and angular diameter of \pp\ can be constrained by 
comparing synthetic spectral energy distributions (SEDs) from the
spherical {\tt PHOENIX} models with the observed spectrophotometry.  
To do so, synthetic photometry is performed using the 13-color
filter sensitivity functions from Johnson \& Mitchell (\cite{johnson2})
compiled by Moro \& Munari (\cite{moro}) and the JHKLM filter 
sensitivity functions from Bessell \& Brett (\cite{bessel}).
The synthetic spectrophotometry values are scaled to the
observed values by the factor
\[ 
\frac{1}{4}\,\theta_{\rm Ross}^2\,10^{-0.4\,\frac{A_{\lambda}}{A_V}\,R_V\,E(B-V)}
\]
where $\theta_{\rm Ross}$ is the wavelength-independent Rosseland
angular diameter, $A_{\lambda}/A_V$ is the extinction curve relative
to the V-band, $E(B-V)$ is the color excess, and we assume $R_V= 3.1 =
A_V/E(B-V)$ using the extinction curves of Cardelli et al. (\cite{cardelli})

A comparison of the best fit synthetic SED to the 13-color photometry,
which covers the widest wavelength range contemporaneously, with all the
spectrophotometric data is shown in Fig.~\ref{fig:sed}.  
The shape of the
near-infrared continuum and the strong TiO bands in the optical
provide the principal constraints on \teff, while the fluxes in
absolute units constrain \adross.  The best fit parameters, based on
least-squares fits to the photometry and application of the F-test
(Aufdenberg et al. \cite{aufdenberg1}) are
\teff\ = 3550$ \pm$ 50 K and \adross\ = 8.0$ \pm$ 0.4 mas.  
Together with the parallax (Table~\ref{tab:psipheprop}), the ranges in
these values translate into the 1- and 2-sigma uncertainty regions on
the theoretical HR diagram in Fig.~\ref{fig:hr}. The uncertainty in the
luminosity is somewhat of an overestimate, since the best fit values
for \teff\ and \adross\ are not independent.

While \teff\ and \adross\ are well constrained by
the SED fit alone, the surface gravity and mass are not.  However,
relying on the evolutionary tracks from Girardi (\cite{girardi}), 
the 2-sigma region on the HR diagram indicates that the mass 
is 1.2 $\pm$ 0.4 \msun. A surface gravity then follows from 
this mass, \adross, and the
parallax: $\log g = 0.66^{+0.18}_{-0.23}$.  Since these constraints are much
tighter than can be derived from comparisons to spectrophotometry
alone, the mass and surface gravity parameters for the synthetic
spectrum shown in Fig.~\ref{fig:sed} were chosen to be consistent with these
values.  Table~\ref{tab:psipheprop} gives a summary of the derived values.
\subsection{ {\tt ATLAS} model atmospheres}
In addition to the {\tt PHOENIX} models, we employ intensity
profiles predicted by standard plane-parallel hydrostatic
{\tt ATLAS\,9} model atmospheres from the Kurucz CD-ROMS 
(Kurucz \cite{kurucz}). 
The Kurucz CDROMs tabulate monochromatic
LD profiles for 17 angles in 1221 frequency intervals
ranging from 9.09\,nm to 160\,000\,nm. In the range of the
near-infrared $K$-band filter (1800-2500\,nm), the frequencies are
sampled in steps of 10\,nm. This data is available for different
chemical abundances, microturbulent velocities, effective temperatures
and surface gravities.
To be consistent, we have chosen the parameters to be closest to those 
determined for our favorite {\tt PHOENIX} model, i.e. the 
model that best fits the spectrophotometry (Sect.~\ref{sec:phoenixspec}).
These parameter values are \teff\ 3500 and 3750\,K, 
$\log g$ 0.5 and 1.0, solar chemical
abundance, and a standard microturbulent velocity of 2\,km\,sec$^{-1}$
(file kurucz.cfa.harvard.edu/grids/gridP00/ip00k2.pck19).

Finally, we employ a plane-parallel hydrostatic 
{\tt ATLAS\,12} model which includes an improved treatment of
the opacities (Kurucz \cite{kurucz12}, \cite{kurucznew}).
We have chosen again the parameters 
of our favorite spherical {\tt PHOENIX} model for reasons
of consistency, i.e. \teff\ =3550\,K, $\log g$=0.7, and solar chemical
abundance. The {\tt ATLAS\,12} intensity profile is tabulated 
at 16 angles for wavelengths from 1.8\,$\mu$m to 2.5\,$\mu$m 
in steps of 0.5\,nm.
\subsection{Effects of model geometry on the interpretation of fitted 
angular diameters}
\label{sec:diaminterp}
We use spherical {\tt PHOENIX} models as well as plane-parallel
{\tt PHOENIX} and {\tt ATLAS} models, as described above, for the fits 
to our interferometric data. The fit result for both geometries is
the LD diameter \adld\ at which the intensity reaches zero
(see Sect.~\ref{sec:broadband} below). 
\begin{figure}
\centering
\resizebox{0.95\hsize}{!}{\includegraphics[angle=0]{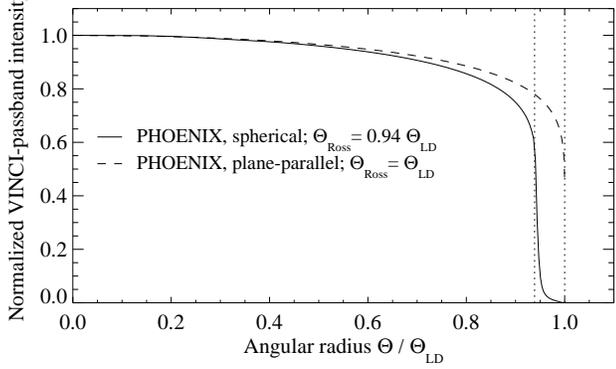}}
\caption{Comparison of CLVs predicted by 
a spherical (solid lines) and a 
plane-parallel (dashed lines) {\tt PHOENIX} model, both with the
same 0\% intensity diameter \adld. 
Both models have parameters \teff\ =3550\,K, $\log g$=0.7.
The spherical model has a mass of 1.3 \msun. Both CLVs are computed
for the broad-band VINCI sensitivity function ($K$-band).
The two CLVs differ by a tail-like extension for the spherical model 
which is missing for the plane-parallel model. 
We describe the diameters of both models by the Rosseland diameter, 
which is close to the point where the CLV drops steeply and most continuum
photons escape. The \adross\ values for these two model 
geometries differ by a factor of 0.94, while the \adld\ values 
correspond. 
}
\label{fig:clv-comp}
\end{figure}
Figure~\ref{fig:clv-comp} illustrates CLV predictions by spherical and
plane-parallel {\tt PHOENIX} models (\teff=3550\,K, $\log g$=0.7,
mass of the spherical model 1.3\,\msun; CLVs are calculated
for the broad VINCI $K$ passband). The monochromatic contributions 
over the $K$-band to this passband-averaged spherical CLV are shown 
below in Fig.~\ref{fig:udld} (middle). It illustrates that
our broad-band measurement is dominated by continuum $K$-band photons,
and that it has only little line contamination (relative to a strong
molecular band). The escaping continuum photons originate 
from a rather compact zone at  $\mu \sim$\,0.3.
The spherical CLV shows an inflection point and a tail-like extension 
which is caused by the optically-thin limb of the spherical model. 
The plane-parallel
model is semi-infinite for all angles $\theta$ and has a singularity
at $\theta=90^\circ$ or $\theta=$\,\adld. This causes the sharp edge of 
the plane-parallel CLV and its lack of any tail-like extension.
As a result, despite the correspondence of the \adld\ values, 
the points where the CLVs drop steeply, which is also the point 
where most continuum photons 
escape, do not correspond for these two model geometries. In other
words, different \adld\ fit values are expected for plane-parallel and
spherical model geometries using the same physical data.
We use the Rosseland diameter \adross, which is associated with a 
Rosseland optical depth of unity, to describe the stellar diameters.
The Rosseland diameter is close to the compact zone where most 
continuum photons escape and where the CLV drops steeply, and is hence 
comparable for both model geometries.\\
For the plane-parallel models, we assume \adld\ $\equiv$ \adross, since
the plane-parallel CLVs drop steeply to zero directly at 
the stellar limb.\\
For the spherical models, \adld\ is associated with
the outermost shell of the model (at $R_0$) and the 
wavelength-independent conversion factor from 
\adld\ to \adross\ is the 
ratio $C_{\rm Ross/LD}=R(\tau_{\rm Ross} = 1)/R_0$.
Here, the outermost radius $R_0$ of the model is defined by
the standard outer boundary conditions for our model atmospheres,
which are a continuum optical depth of 1e-6 at 1.2\,$\mu$m and 
an outer gas pressure of 1e-4\,dynes/cm$^2$. 
The inner optical depth boundary is 100 at 1.2\,$\mu$m.
The radial grid and outermost radius differ from model to model, 
resulting in a model-dependent conversion factor $C_{\rm Ross/LD}$
(as used in Table~\ref{tab:fit}, Sect.~\ref{sec:comp} below).
\section{VLTI/VINCI measurements}
\subsection{Observations}
\label{sec:observations}
The ESO VLTI, located on Cerro Paranal in northern Chile, 
is currently in the phase of commissioning. For a recent general
overview see Glindemann et al. (\cite{glindemann}) and references
therein. First fringes using 40\,cm test siderostats and the 
commissioning instrument VINCI (Kervella et al. \cite{kervellavinci2})
were achieved in March 2001, 
followed by first fringes using the 8.2\,m diameter Unit Telescopes (UTs) 
on October 30, 2001. 
First fringes with the first scientific instrument, the mid-infrared
instrument MIDI (Leinert et al. \cite{midi}), were obtained in December 2002.

Our observations of \pp's squared visibility amplitudes beyond the first
minimum employing the UTs succeeded one month after the achievement of
first fringes with the UTs, and were first mentioned in ESO press release
23/01 (Nov. 5, 2001).
These visibility values of \pp\ were obtained during the nights starting 
on Nov. 1\,\&\,2, 2001 using the UT\,1 -- UT\,3 102\,m baseline of the 
VLTI, and the VINCI instrument, 
as part of the VLTI commissioning. The UTs were
not yet equipped with adaptive optics nor a tip-tilt corrector.
High-contrast fringes at short spatial frequencies were obtained before 
and after the UT measurements using the 40\,cm test siderostats.
These data, as all scientifically interesting VLTI/VINCI data, 
have been made publicly available through the ESO archive. 
Table~\ref{tab:obslog} details our observations and 
Table~\ref{tab:calibrators} shows the adopted properties of the 
used calibration stars.

The use of the large apertures of the 8\,m UTs together with the 
use of single-mode optical fibers, a technique that is known to 
lead to very accurate visibility measurements 
(see, e.g. Coud\'e du Foresto \cite{coude98b}),  
has enabled precise measurements of low visibility amplitudes.
Bootstrapping techniques, such as those used by 
Hajian et al. (\cite{hajian}) and Wittkowski et al. (\cite{wittkowski1}) 
to detect and track fringes with weak contrasts, were not necessary. 

Configurations using different aperture sizes for different baseline
lengths, as used here, might be an interesting perspective for the 
completed VLTI as well. The completed VLTI will routinely allow
the measurement of the limb-darkening effect for many stars, and open the
possibility to obtain stellar surface structure parameters beyond
limb-darkening (see e.g., Wittkowski et al. \cite{wittkowski2}).
This will require visibility data uniformly distributed over the 
$uv$-plane including both, high-contrast fringes in the first lobe
as well as very 
low-contrast fringes in the second and following lobes
of the visibility function. Then, the lowest fringe contrasts can be 
measured with the 8\,m UTs in order to obtain
a sufficient signal-to-noise ratio (S/N) for fringe detection and tracking 
and/or an enhanced S/N for these important low visibility points while 
higher fringe contrasts on the same source can be obtained 
with the 1.8\,m ATs.
\begin{table}
\centering
\caption{Record of our \pp\ observations. Listed are the date of 
observation, the used stations, the baseline length, the 
number of series of interferograms, and the used 
calibration stars.\label{tab:obslog}}
\begin{tabular}{llrrl}
Date & Stations & $B$ (m) & \#  & Calibration stars \\ \hline
2001-10-12 & E0/G0 & 16 & 4 & \object{$\beta$\,Cet}, \object{$\epsilon$\,Lep} \\    
2001-10-16 & E0/G0 & 16 & 3 & $\beta$\,Cet, \object{$\eta$\,Cet} \\ 
2001-10-17 & E0/G0 & 16 & 3 & $\beta$\,Cet, $\eta$\,Cet \\
2001-10-18 & E0/G0 & 16 & 3 & $\beta$\,Cet, $\eta$\,Cet \\
2001-11-01 & UT\,1/UT\,3 & 102 & 13 & \object{$\chi$\,Phe}, \object{$\gamma^2$\,Vol} \\
2001-11-02 & UT\,1/UT\,3 & 102 & 5 & \object{$39$\,Eri}, $\gamma^2$\,Vol \\
2001-11-04 & E0/G0 & 16        & 7 & $\beta$\,Cet, $\epsilon$\,Lep \\
2001-11-05 & E0/G0 & 16        & 3 & $\beta$\,Cet, $\epsilon$\,Lep \\
\end{tabular}
\end{table}
\begin{table}
\centering
\caption{Properties of the used calibration stars. 
Listed are the spectral type from Perryman \& ESA (\cite{esa}), 
the $K$-band magnitude from Gezari (\cite{gezari}), 
the effective temperature \teff\ derived from the spectral type, 
the effective wavelength $\lambda_0$ derived from \teff\ and the 
VINCI sensitivity
curve, the adopted $K$-band UD diameter and its error $e\theta$.
The sources for the adopted diameter values and their errors are
(1) Cohen et al. (\cite{cohen}) and/or
(2) Bord\'e et al. (\cite{borde}).
\label{tab:calibrators}}
\begin{tabular}{lrrcrrrrl}
Obj. & SpT & K & $T_{\rm eff}$ & $\lambda_{\rm eff}$ & $\theta_{\rm UD}^K$ & 
$e\theta$ & Ref. \\
  & &  & K              & $\mu$m & mas & mas & \\ \hline
$\beta$\,Cet    & K0III  & -0.2 & 4800 & 2.179 & 5.18& 0.06 & 1,2 \\
$\epsilon$\,Lep & K4III  & -0.2 & 4075 & 2.181 & 5.91& 0.06 & 1,2 \\
$\eta$\,Cet     & K2III  & 0.9 & 4600 & 2.180 & 3.35& 0.04 & 1,2 \\
$\chi$\,Phe     & K5III  & 1.5 & 4000 & 2.181 & 2.69& 0.03 & 1,2 \\
$\gamma^2$\,Vol & K0III  & 1.5$^a$ & 4800 & 2.179 & 2.44 & 0.06 & 1\\
$39$\,Eri       & K3III  & 2.3 & 4200 & 2.181 & 1.81& 0.02 & 1\\
\end{tabular}\\
$^a$ from $V$ and the $V-K$ color for a K0\,III star.
\end{table}

Our data were recorded through the VINCI $K$-band filter. The scan length of the optical path difference was
280\,$\mu$m, the fringe frequency, i.e. the time needed to scan one 
interferometric fringe, was 296\,Hz for the siderostat data and 695\,Hz for
the UT data. Our data were taken as series of 100 or 500 interferogram scans.

\subsection{Data reduction}
\label{sec:reduction}
\begin{figure}
\centering
\resizebox{1.0\hsize}{!}{\includegraphics{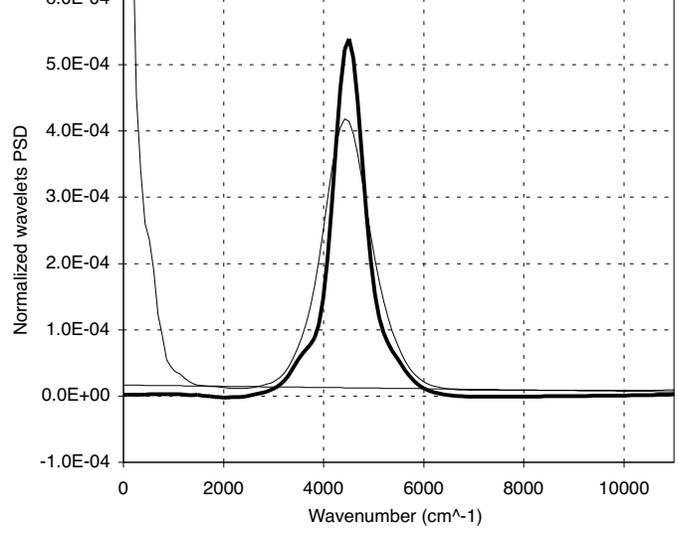}}
\caption{Average wavelets PSD of one
series (Date 2001-11-02, Univ. Time 05:23:23) of 414 accepted 
interferograms obtained on \pp\ (UT1-UT3 baseline). 
The thin curve is the simple
average of all PSD before removal of the background. The background
is estimated by a fit of a linear function to low and high frequency 
PSD values outside the fringe signal (ranges
[1400..2000]\,cm$^{-1}$ and [8000..9988]\,cm$^{-1}$) and compensated.
The squared coherence factor is derived from the integration 
of the background calibrated PSD (thick line) in the range 
[2000..8000]\,cm$^{-1}$.
Before averaging, each fringe peak is re-centered to the 
theoretical frequency of the fringes in order to reduce the energy 
spreading due to the differential piston effect before the power peak 
integration. 
This is the reason why the shape of the 
non-re-centered (thin line) and re-centered and bias corrected 
(thick line) PSDs are not exactly the same.
}
\label{fig:bias}
\end{figure} 
We have used a modified version (Kervella et al. \cite{kervella}) of the 
standard VINCI data reduction pipeline (Ballester et al. \cite{ballester}), 
whose general principle is based on the original algorithm
of the FLUOR instrument (Coud\'e du Foresto et al. \cite{cdf97};
Coud\'e du Foresto et al. \cite{coude98a}). 
This data reduction software includes the following procedures.

For each scan, the two interferograms produced from the two 
interferometric outputs of the VINCI beam combiner are calibrated
for photometric intensity fluctuations using the two
photometric outputs. After this first photometric calibration, 
the two calibrated interferograms are subtracted to remove residual
photometric fluctuations. Since the two fringe patterns are in 
perfect phase opposition, this subtraction removes a large part of the 
correlated fluctuations while enhancing the interferometric fringes.
This is particularly important for the small \pp\ visibility amplitudes
in the second lobe of the visibility function obtained with the 
UT1-UT3 baseline. 
Instead of the classical Fourier analysis, the 
time-frequency analysis (S\'egransan et al. \cite{s99})
based on the continuous wavelet transform (Farge \cite{farge92}) is used
to derive the wavelet power spectral density (wavelet PSD) of each scan.
The differential piston corrupts the amplitude and the shape of the
fringe peak in the PSD.
A selection of single scans 
is used to remove interferograms that are affected by strong differential
piston and, if existent, that show no fringes. 
The selection criteria include checks of
the peak width in the time domain ($\pm 50$\% around theoretical 
value is accepted), as well as the peak position 
($\pm 30$\% around theoretical value is accepted) and peak width in the 
frequency domain ($\pm 40$\% around theoretical 
value is accepted). In addition, fringe scans with low photometric 
signal (S/N of less than 5), with a large jump of the optical path 
difference before the scan (larger than 20\,$\mu$m), and
with fringes at the edge of the scan are rejected.
The average wavelet PSD of the selected interferograms is computed.

The use of squared quantities, such as the PSD, requires attention to 
a possible additive background of the PSD caused by residual photon and 
detection noise, despite the first compensations mentioned above.
The relative contribution of remaining biases would be 
highest for small visibility values, such as those studied here 
(cf. Wittkowski et al. \cite{wittkowski1}).
In the case of the VINCI data reduction, the residual noise background
is determined from a least squares fit to low and high 
frequency PSD values outside the fringe signal, and compensated.
Figure~\ref{fig:bias} shows an example of the average wavelet PSD 
derived from one series of 414 accepted interferograms obtained on \pp\ 
using UT1 and UT3. Shown are the
raw PSD before compensation of the residual PSD background, 
the derived model of the PSD background, and the compensated PSD.
Despite the very strong photometric fluctuations
that are observed in the multi-speckle regime, and despite 
the very low fringe signal
(for this particular PSD, the squared visibility amplitude of \pp\ 
was only $|V|^2 = 0.013$), the background calibrated 
wavelet PSD is free of any photometric contamination. This is very important
to reliably obtain our small \pp\ squared visibility values beyond the 
first minimum.

The coherence factor values are finally derived by integrating 
the compensated average wavelet PSD of the selected 
interferograms. This integration includes fringe power at all 
wavelengths over the broad VINCI sensitivity band.
Single scans with coherence factors that differ by more than 
3 $\sigma$ from the median of the sample are rejected.
\subsection{Calibration}
\label{sec:calibration}
Table~\ref{tab:obslog} lists the calibration stars that have been used during
the different nights of our \pp\ observations; 
Table~\ref{tab:calibrators} details their characteristics
including the adopted diameters and their errors.

Owing to the low limiting magnitude of the siderostats 
($K_{\rm corr}\,\simeq\,1$ at the time of these observations), 
the calibration stars for the 16\,m siderostat baseline have diameters, 
between 3\,mas and 6\,mas, which are not small compared to the 
expected \pp\ diameter of $\simeq$\,8\,mas. Thus, the calibration 
requires special attention to the adopted diameter values.
Cohen et al. (\cite{cohen}) have used a spectro-photometric calibration
to derive high-precision Rosseland diameters and their errors 
for a list of 422 sources. They found that their estimates compare
well with several other diameter estimates. This was confirmed by
Bord\'e et al. (\cite{borde}), who in addition reduced this list
to 374 stars carefully selected to be used as calibration stars for
long baseline stellar interferometry.
We rely on these diameter estimates and their errors as given in 
the lists by Cohen et al. (\cite{cohen}) and Bord\'e et al. (\cite{borde}).
The main calibration star for the siderostat 
data, $\beta$\,Cet, used during all siderostat nights, was in 
August 2002 calibrated against the Cohen stars 20\,Cet 
(M0\,III, $\theta_{\rm UD}^K=3.41 \pm 0.037$\,mas) and 
$\iota$\,Cet (K1\,III, $\theta_{\rm UD}^K=3.27 \pm 0.036$\,mas)
using the 64\,m G1--E0 baseline. The measured $\beta$\,Cet diameter was
found to be well consistent with the Cohen value and its error given 
in Table~\ref{tab:calibrators}. An additional calibration uncertainty
arises from the fact that the 16\,m baseline measurements, 
corresponding to squared visibility amplitudes in the range $\sim$\,0.8-0.95, 
only marginally resolve our
object \pp. To take these calibration uncertainties for the 16\,m baseline
into account, we weight the measurements obtained with 
this baselines by a factor of two lower than the 102\,m baseline 
measurements.
A VLTI program in order to derive calibration star diameters 
in a self-consistent way from VLTI observations is in 
progress (Percheron et al. \cite{percheron}).

For the \pp\ data sets as well as all calibration star data sets, 
the coherence factors $c^2$ are computed as described 
in Sect.~\ref{sec:reduction}.
The transfer function $t^2$ is computed for all calibration star data sets as
$t^2=c^2/|V|^2$, where $|V|^2$ is the adopted squared visibility amplitude
for the respective time and baseline using the adopted diameter value 
and effective wavelength from Table~\ref{tab:calibrators}. 
The transfer function value for the time
of each of the \pp\ observations is then obtained as a weighted
average of the calibration stars' $t^2$ values during the night 
with the weight 
being a product of (1) a Gaussian function of the time difference 
from the respective \pp\ measurement with a width $\sigma$ of 3 hours for the
siderostat data and 2 hours for the UT data, and (2) the error of the 
single transfer function measurement consisting of the statistical 
$c^2$ error and the error resulting from the adopted diameter uncertainty.
The final \pp\ squared visibility values are then obtained by division 
of the \pp\ coherence factors by the transfer function values obtained 
for the respective time.
The final errors of the \pp\ squared visibility amplitudes are
computed from the error of the \pp\ coherence factor (the scatter of the 
single scan's coherence factors) and the uncertainty of 
the transfer function values, which include the errors of the 
calibration stars' coherence factors, the adopted diameter errors, 
and the variation of the computed transfer function over the night. 
\subsection{Results}
\begin{table}
\centering
\caption{Resulting squared visibility amplitudes $|V|^2$ for each series
of interferograms.
Given are the date, the Universal Time,
the spatial frequency $B/\lambda_0$ (1/\arcsec) for \pp's effective 
wavelength of $\lambda_0$=2.183\,$\mu$m, the squared 
visibility amplitudes, their error $\sigma(|V|^2)$, and the 
number (\#) of accepted scans.}
\label{tab:results}
\begin{tabular}{llrrrr}
Date       & Univ. Time   & $B/\lambda_0$    & $|V|^2$ & $\sigma$ & \# \\
2001-      &          &  (1/\arcsec)  &         &          &    \\\hline
10-13 & 03:40:05 & 35.29 & 8.471e-01 & 4.902e-02 & 462 \\
10-13 & 04:24:29 & 35.55 & 8.580e-01 & 5.038e-02 & 439 \\
10-13 & 04:59:41 & 35.39 & 8.084e-01 & 4.938e-02 & 460 \\
10-13 & 05:41:03 & 34.74 & 8.306e-01 & 5.054e-02 & 452 \\
10-17 & 01:56:26 & 33.82 & 8.805e-01 & 2.306e-02 & 479 \\
10-17 & 04:48:29 & 35.34 & 8.721e-01 & 2.650e-02 & 450 \\
10-17 & 06:13:42 & 33.29 & 8.739e-01 & 2.742e-02 & 483 \\
10-18 & 01:54:08 & 33.86 & 8.666e-01 & 2.082e-02 & 463 \\
10-18 & 03:12:03 & 35.20 & 8.409e-01 & 2.270e-02 & 395 \\
10-18 & 04:35:33 & 35.43 & 8.626e-01 & 2.442e-02 & 433 \\
10-19 & 03:02:01 & 35.12 & 8.694e-01 & 1.754e-02 & 444 \\
10-19 & 04:09:31 & 35.54 & 8.668e-01 & 1.712e-02 & 481 \\
10-19 & 05:25:09 & 34.55 & 8.625e-01 & 1.739e-02 & 464 \\
11-02 & 05:17:37 & 201.09 & 1.196e-02 & 7.545e-04 & 50 \\
11-02 & 05:20:05 & 200.53 & 1.318e-02 & 7.164e-04 & 398 \\
11-02 & 05:23:23 & 199.76 & 1.317e-02 & 7.220e-04 & 414 \\
11-02 & 05:25:21 & 199.30 & 1.325e-02 & 7.388e-04 & 62 \\
11-02 & 05:27:49 & 198.70 & 1.318e-02 & 7.309e-04 & 413 \\
11-02 & 05:31:00 & 197.93 & 1.317e-02 & 7.371e-04 & 411 \\
11-02 & 05:34:13 & 197.13 & 1.317e-02 & 7.425e-04 & 433 \\
11-02 & 06:20:40 & 183.95 & 9.144e-03 & 7.091e-04 & 66 \\
11-02 & 06:32:42 & 180.01 & 8.605e-03 & 7.322e-04 & 77 \\
11-02 & 06:35:13 & 179.16 & 8.391e-03 & 7.009e-04 & 308 \\
11-02 & 06:38:38 & 178.00 & 8.000e-03 & 6.891e-04 & 329 \\
11-02 & 06:42:02 & 176.82 & 7.847e-03 & 6.887e-04 & 286 \\
11-03 & 02:50:05 & 221.37 & 1.266e-02 & 1.385e-03 & 81 \\
11-03 & 02:52:31 & 221.20 & 1.197e-02 & 1.360e-03 & 387 \\
11-03 & 02:55:47 & 220.96 & 1.180e-02 & 1.369e-03 & 354 \\
11-03 & 02:58:52 & 220.72 & 1.221e-02 & 1.410e-03 & 386 \\
11-03 & 03:02:10 & 220.46 & 1.223e-02 & 1.430e-03 & 426 \\
11-05 & 04:56:32 & 33.37 & 8.622e-01 & 2.822e-02 & 348 \\
11-05 & 05:06:35 & 32.98 & 8.767e-01 & 2.458e-02 & 492 \\
11-05 & 05:11:46 & 32.77 & 8.804e-01 & 2.466e-02 & 492 \\
11-05 & 06:13:25 & 29.69 & 8.832e-01 & 2.218e-02 & 195 \\
11-05 & 06:25:05 & 29.00 & 9.048e-01 & 2.592e-02 & 413 \\
11-05 & 06:35:29 & 28.36 & 9.159e-01 & 2.572e-02 & 453 \\
11-05 & 07:53:25 & 23.19 & 9.643e-01 & 2.640e-02 & 427 \\
11-06 & 01:41:53 & 35.00 & 8.640e-01 & 2.294e-02 & 454 \\
11-06 & 01:46:19 & 35.06 & 8.598e-01 & 2.312e-02 & 484 \\
11-06 & 01:50:41 & 35.12 & 8.612e-01 & 2.336e-02 & 479 \\
\end{tabular}
\end{table}
Table~\ref{tab:results} shows the resulting 
squared visibility amplitudes $|V|^2$
for each of our data sets. The given errors of $|V|^2$
include  the statistical error obtained from the scatter
of the single scan values, the uncertainty of the calibration star's
diameter (see Table~\ref{tab:calibrators}), and the variation of the
computed transfer function over the night
(see Sect.~\ref{sec:calibration}).
A plot (Figure~\ref{fig:vis}) of our measured $|V|^2$ values together with
best fitting {\tt PHOENIX} and {\tt ATLAS} models is discussed in
Sect.~\ref{sec:comp} (below).
\subsection{Computation of broad-band model visibility values}
\label{sec:broadband}
\begin{figure}
\centering
\resizebox{1.0\hsize}{!}{\includegraphics[]{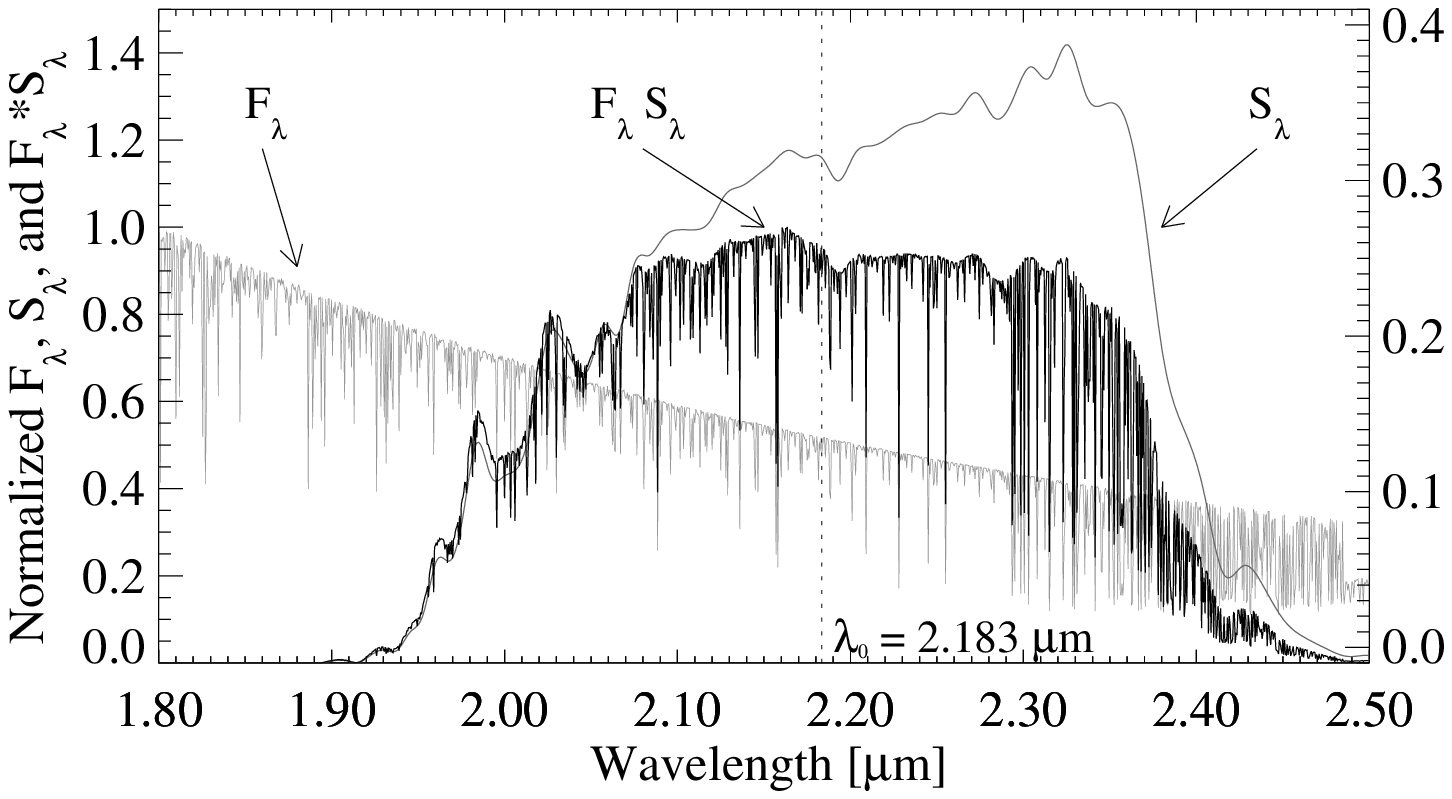}}

\resizebox{1.0\hsize}{!}{\includegraphics[]{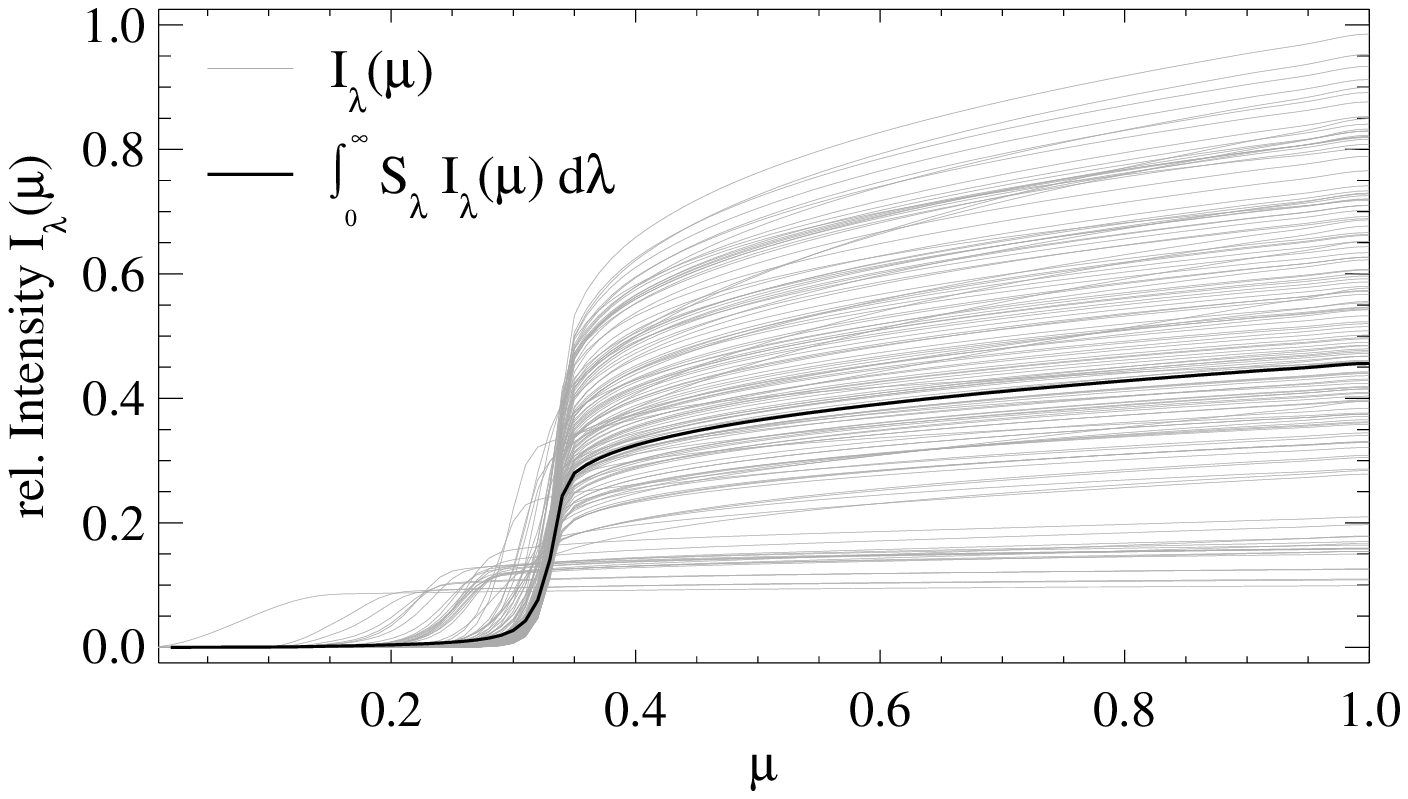}}

\resizebox{1.0\hsize}{!}{\includegraphics[]{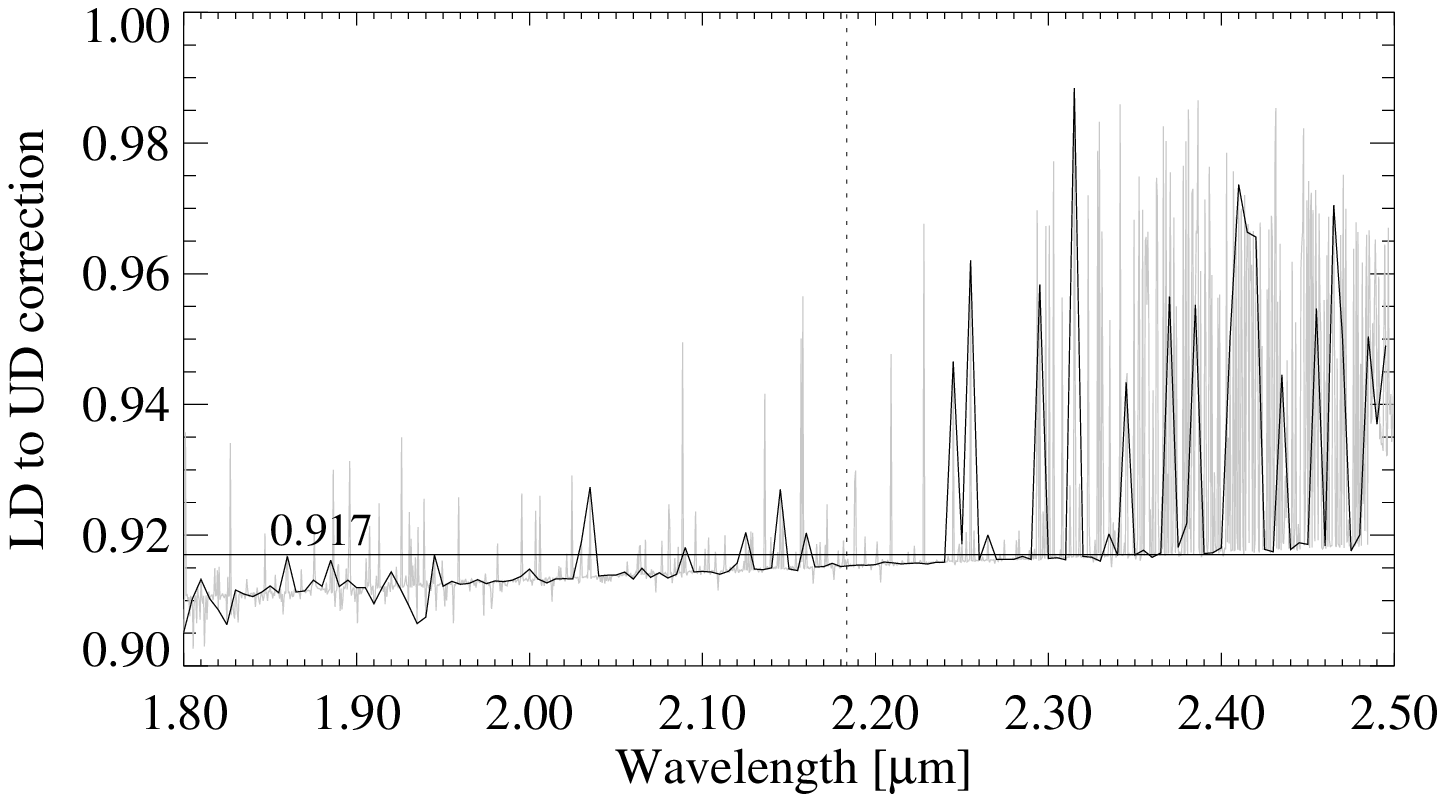}}

\caption{(Top) Model \pp\ $K$-band spectrum $F_\lambda$ (gray), 
VINCI sensitivity function $S_\lambda$ (gray; right scale), and effective
spectrum $F_\lambda\,S_\lambda$ (black). The dashed vertical line
denotes the effective wavelength $\lambda_0=2.183\,\mu$m.
(Middle) Monochromatic model \pp\ intensity profiles across the 
$K$-band with a 
resolution of 50\,nm (gray) and their $S_\lambda$ weighted average 
(calculated with 5\,nm resolution, black). 
With VLTI and AMBER, it will
in principle be possible to directly measure these different intensity
profiles with a resolution of up to $\sim$ 0.2\,nm.
(Bottom) Monochromatic correction factors 
from LD to UD diameters with 
a resolution of 5\,nm (gray) and 50\,nm (black) between 1.8\,$\mu$m and
2.5\,$\mu$m. The horizontal line
denotes the weighted average of the correction factors (value 0.917).
The spherical {\tt PHOENIX} model as used 
in Fig.~\protect\ref{fig:sed} was employed for
these plots. The spectrum shows CO lines at the red end, which
have a significant impact on the \pp\ intensity profiles at 
{\it these} wavelengths.
They become optically thin near the limb and hence, the limb 
is less dark, the 
radius is larger in the bands, and the LD-to-UD correction factor
is closer to unity. At most wavelengths the limb edge is at $\mu\sim 0.3$, 
but at a minority of wavelengths the limb is brighter, 
almost uniform at one wavelength. These profiles come from wavelengths 
in the CO band. These CO lines, however, do not have a strong
influence on the VINCI passband-averaged model intensity profile
because of their narrowness, their lower intensity, and the 
shape of $S_\lambda$.
The LD-to-UD correction factors show also the general trend 
that the (continuum) intensity profiles become closer to a 
uniform disk with increasing wavelength.
}
\label{fig:udld}
\end{figure} 
The use of the VINCI broad-band $K$ filter influences the 
obtained squared visibility amplitudes in two ways 
(cf. Tango \& Davis {\cite{tango}). Firstly,
the stellar intensity profile is not observed at a monochromatic
wavelength $\lambda$ but integrated over the broad-band 
sensitivity curve. The strength of the limb-darkening effect for \pp, 
is expected to vary over the VINCI sensitivity curve. 
Figure~\ref{fig:udld} shows the
expected variation of \pp's flux, intensity profile, and 
strength of the limb-darkening (correction factors from LD 
to UD diameters) over the VINCI passband. The VINCI 
passband-averaged intensity profile and strength of the limb-darkening 
are shown as well. The plots show that our $K$-band measurements are
dominated by continuum photons and that there is only little line
contamination (relative to a strong molecular band).
Secondly, squared visibility amplitudes are averaged over different spatial
frequencies ($B/[(\lambda_0-\Delta\,\lambda)..(\lambda_0+\Delta\,\lambda)]$) 
for any point of observation with fixed
baseline $B$ and a broad-band filter (central wavelength $\lambda_0$, 
filter width $\Delta\,\lambda$) . This latter effect has been discussed by 
Kervella et al. (\cite{kervella}) for the VINCI instrument and been
referred to as ``bandwidth smearing''. 

The calculation of the synthetic broad-band squared visibility amplitudes 
has to follow the way the measured squared visibility amplitudes are 
derived from the raw fringe data. 
For the VINCI data reduction (see Sect.~\ref{sec:reduction}), first 
the PSD, a squared quantity corresponding to the squared visibility
amplitudes, is computed from the interferograms, 
and then the coherence factor is derived from the integral of the PSD,
which includes fringe power from all wavelengths over the sensitivity band. 
Hence, for VINCI, the model squared visibility 
amplitudes at different wavelengths and spatial frequencies have to be 
integrated (see Kervella et al. \cite{kervella}),
and not the complex visibilities (visibility amplitudes with sign) 
as implied by the general formalism in Tango \& Davis (\cite{tango}).
This means that the broad-band squared visibility amplitude 
values never reach zero in our case.

We derive non-normalized monochromatic LD model visibility values 
$V_{\rm LD}(\lambda)$ on a grid of wavelengths 
(range 1.8..2.5\,$\mu$m in steps of 0.5\,nm for our {\tt PHOENIX} models) by 
numerical evaluation of the Hankel transform 
of the intensity profiles $I_\lambda^\mu$ tabulated 
for each of these (1400) monochromatic wavelengths $\lambda$ 
(following Davis et al. \cite{davis} and Tango \& Davis \cite{tango},
see also the use in Wittkowski et al. \cite{wittkowski1}) by
\begin{equation}
V_{\rm LD}(\lambda)=
\int_0^1\,S_\lambda\,I_{\lambda}^\mu\,J_0[\pi\,\theta_{\rm LD}\,(B/\lambda)\,(1-\mu^2)^{1/2}]\,\mu\,d\mu
\label{eq:ld}
\end{equation}
Here, $I_\lambda^\mu$ is used as a function of $\mu$, 
$\mu=\cos\theta$ ($\mu \in [0,1]$, $\mu$=0 corresponds to the star's 
radius at which the intensity reaches zero, and $\mu$=1 to the star's center);
$S_\lambda$ is the VINCI instrument's sensitivity curve including
the transmissions of the atmosphere, the optical fibers, 
the VINCI $K$-band filter, and the detector quantum efficiency;
$B$ is the sky-projected baseline length.
For reasons of numerical accuracy, it had been necessary to interpolate 
the tabulated intensity profiles onto a regularly sampled 
grid of $\mu$ values (we have chosen a cubic-spline interpolation 
onto 100 $\mu$ values), rather than interpolating the more complex 
argument of the integral in Eq.~\ref{eq:ld}. 
The integral is then computed using a 5-point Newton-Cotes algorithm. 
We have checked the accuracy of the algorithm by comparing results for 
tabulated UD and FDD intensity profiles with 
their available analytic expressions for $V$. 
In order to calculate a spatial frequency $B/\lambda_0$, as used in
Fig.~\ref{fig:vis} (below), an effective wavelength $\lambda_0$ is 
computed as 
\begin{equation}
\lambda_0=\frac{\int_0^\infty\,S_\lambda\,F_\lambda\,\lambda\,d\lambda}{\int_0^\infty\,S_\lambda\,F_\lambda\,d\lambda}
\end{equation}
with $F_\lambda$ being the stellar flux at wavelength $\lambda$:
\begin{equation}
F_\lambda=\int_0^1\,I_\lambda(\mu)\,\mu\,d\mu.
\end{equation}
With our favorite spherical {\tt PHOENIX} model for \pp\ as derived in 
Sect.~\ref{sec:phoenixspec}, Fig.~\ref{fig:sed}, 
we obtain $\lambda_0=2.183\,\mu$m. This value characterizes the wavelength
of our measurement, but is not used for the calculation 
of the synthetic visibilities since they are calculated for the full 
wavelength range in Eq.~\ref{eq:ld}.
The functions $S_\lambda$ and $F_\lambda$, as well as the 
effective wavelength are shown in Fig.~\ref{fig:udld} (top).
As for the processing of the instrument data, each monochromatic 
visibility is not normalized separately.
The VINCI passband-averaged squared visibility amplitudes are finally 
computed as
\begin{equation}
|V_{\rm LD}|^2=\frac{\int_0^\infty\,|V_{\rm LD}(\lambda)|^2\,d\lambda}
{\int_0^\infty S_\lambda^2\,F_\lambda^2\,d\lambda},
\end{equation}
including the proper normalization by the total detected flux.

We directly use the model atmosphere's tabulated intensity profiles
(the stellar radiation field)
in Eq.~\ref{eq:ld}. No approximation of the intensity profile 
by any limb-darkening law is used. This ensures two important points:
Parametrizations of intensity profiles are usually
very good approximations for the visibility curve before the first minimum, 
but may lead to deviations from the original visibility profile beyond 
the first minimum, which may be significant with the precision of our 
measurement.
Furthermore, the direct use of the tabulated model stellar radiation field
on a fine grid of wavelengths enables us to compute the 
model squared visibility values for exactly our instrument's passband, 
while published limb-darkening coefficients are calculated for specific
filter curves which do not well resemble the VINCI sensitivity curve
including e.g. the transmission of the optical fibers and the detector
efficiency. 

As discussed in Sect.~\ref{sec:diaminterp}, the LD diameter used 
in Eq.~\ref{eq:ld}
depends on the detailed model structure 
and is therefore not a well suited quantity for reference.
We transform \adld\ into the Rosseland
angular diameter \adross\ with the 
factor $C_{\rm Ross/LD}$ defined in Sect.~\ref{sec:diaminterp}. 
\section{Comparison of our VLTI/VINCI data to our model predictions}
\label{sec:comp}
We use the different model atmospheres as described 
in Sect.~\ref{sec:models} and, with the formalism described above
in Sect.~\ref{sec:broadband}, calculate synthetic squared visibility 
amplitudes for the spatial frequencies of our 
data. For each considered
model, we find by a non-linear least squares fit the \adld\ value
for which the synthetic and measured squared 
visibility amplitudes have the lowest $\chi^2$ value.
The LD angular diameter \adld\ is treated as 
the only free parameter.
The best fitting \adld\ value is then transformed into
the Rosseland angular diameter \adross\ using the factors 
$C_{\rm Ross/LD}$ defined in Sect.~\ref{sec:diaminterp}.

Table~\ref{tab:fit} lists the fit results for the different
spherical {\tt PHOENIX}, and plane-parallel {\tt PHOENIX},
{\tt ATLAS\,12}, and {\tt ATLAS\,9} models.
The first three columns specify the model parameters \teff, $\log g$,
and $M$, followed by the model-specific factor $C_{\rm Ross/LD}$.
The next two columns give the best fitting \adld\ values together with
the reduced $\chi^2_\nu$ values. Finally, the results for \adross\ are 
given in the last column.
The formal errors for \adld\ are found to be about uniform for all
considered models. The standard formal error $\sigma(\theta_{\rm LD})$
for \adld, derived
as the $\Delta\theta_{\rm LD}$ corresponding to $\Delta\chi^2=1$, 
is $\sim$\,0.05\,mas. Deriving the error by means of the $F$-test,
which is a more reliable estimate than the $\chi^2$-test
(e.g., Bevington \& Robinson \cite{bevington}), we obtain  
$\sigma(\theta_{\rm LD})\sim$\,0.12\,mas. Here, we assume a number of 
degrees of freedom of 39, i.e. the number of visibility points minus one. 
The real situation is more complicated because (1) the errors 
include statistical errors as well as 
systematic errors (which can not easily be separated), (2) 
some squared visibility value measurements are not independent but 
linked via the same calibration measurements, (3) the visibility function 
is sampled at two different spatial frequency regions using different
telescopes, 
and (4) obtained $\chi^2_\nu$ values are larger than unity. 
We estimate the final error $\sigma(\theta_{\rm LD})$, including calibration 
uncertainties, to be $\sim$\,0.2 mas by comparing results obtained by
different calibrations and different sub-samples of our data.
\begin{table}
\centering
\caption{Results for \adld\
obtained by fits to our interferometric data. Listed are the model input 
parameters \teff, $\log g$, and mass $M$, the model-specific 
correction factors $C_{\rm Ross/LD}$, the fit results for
\adld, their corresponding reduced $\chi^2_\nu$ values, and finally
the Rosseland angular diameter \adross. The 
error $\sigma(\theta_{\rm LD})$ is estimated to be 
uniformly $\pm$ 0.2\,mas for all model fits (see text).}
\label{tab:fit}
\begin{tabular}{llll|ll|l}
\teff & $\log g$ & $M$  &  $C_{\rm Ross/LD}$  & \adld\ & $\chi^2_\nu$ & \adross \\[1ex] \hline\\
\multicolumn{7}{l}{\tt Spherical PHOENIX models:} \\[1ex]
3550  & 0.7      & 1.3  &  0.9388 & 8.664 & 1.80 & 8.13 \\[2ex]
3500  & 0.7      & 1.3  &  0.9394 & 8.663 & 1.81 & 8.14 \\
3600  & 0.7      & 1.3  &  0.9381 & 8.665 & 1.79 & 8.13 \\
3550  & 0.5      & 1.3  &  0.9218 & 8.814 & 1.79 & 8.12 \\
3550  & 1.0      & 1.3  &  0.9577 & 8.504 & 1.79 & 8.14 \\
3550  & 0.7      & 1.0  &  0.9302 & 8.739 & 1.80 & 8.13 \\[1ex]
\multicolumn{7}{l}{\tt Plane-parallel PHOENIX  model:} \\[1ex]
3550  & 0.7      & /    &  1      & 8.168  & 1.72 & 8.17 \\[1ex]
\multicolumn{7}{l}{\tt Plane-parallel ATLAS\,12 model:} \\[1ex]
3550  & 0.7      & /    &  1      & 8.191  & 1.78 & 8.19 \\[1ex]
\multicolumn{7}{l}{\tt Plane-parallel ATLAS\,9 models:} \\[1ex]
3500  & 0.5      & /    &  1      & 8.244  & 1.74 & 8.24 \\
3500  & 1.0      & /    &  1      & 8.243  & 1.73 & 8.24 \\
3750  & 0.5      & /    &  1      & 8.227  & 1.71 & 8.23 \\
3750  & 1.0      & /    &  1      & 8.228  & 1.71 & 8.23 \\
\end{tabular}
\end{table}
Table~\ref{tab:fit} shows results for our favorite model atmosphere from 
Sect.~\ref{sec:phoenixspec}, i.e. the spherical {\tt PHOENIX}
model with \teff=3550\,K, $\log g$=0.7, and $M$=1.3\msun.
In addition, we consider spherical {\tt PHOENIX} models with varied 
parameters that are still consistent with the estimates in 
Sect.~\ref{sec:phoenixspec}. Furthermore, we compare 
plane-parallel {\tt PHOENIX}, {\tt ATLAS\,12}, and {\tt ATLAS\,9} models 
with, for reasons of consistency, parameters closest to 
our favorite model.
 
Figure~\ref{fig:vis} plots our measured squared visibility values
together with the model prediction by our favorite 
spherical {\tt PHOENIX} model with best fitting \adld\ value. 
Shown are also the predictions by corresponding 
plane-parallel {\tt PHOENIX}, {\tt ATLAS\,12}, and {\tt ATLAS\,9} models. 
As a reference of the strength of the limb-darkening, 
UD ($I=1$ for $0\le\mu\le 1$ and $I=0$ otherwise) and 
FDD ($I=\mu$) model visibility functions are shown, with 
diameters $\theta_{\rm UD}$ and $\theta_{\rm FDD}$ corresponding to our 
favorite {\tt PHOENIX} model fit. Our measurements differ significantly
from UD and FDD models, confirming the
limb-darkening effect.
\begin{figure*}
\centering
\resizebox{0.48\hsize}{!}{\includegraphics{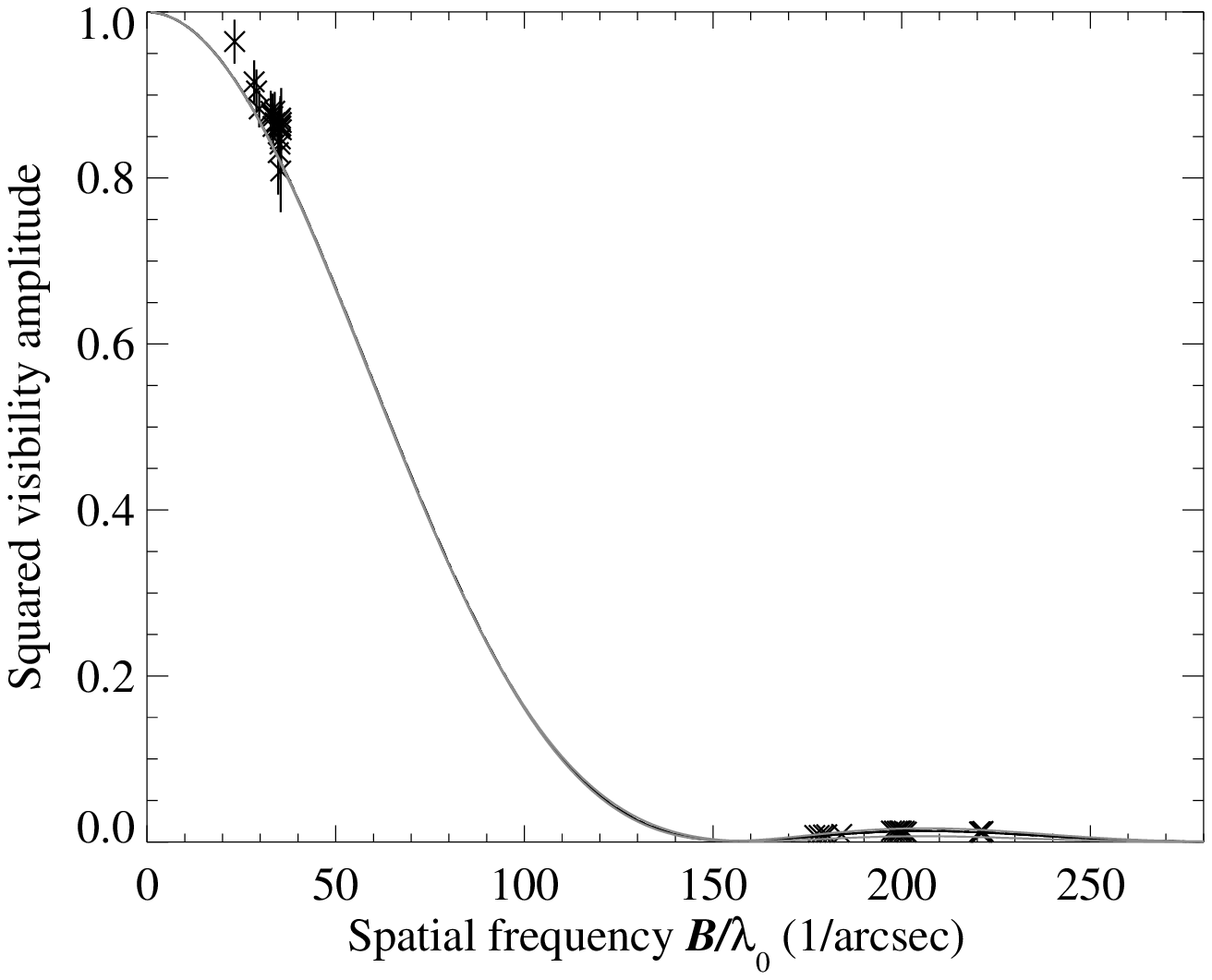}}
\resizebox{0.48\hsize}{!}{\includegraphics{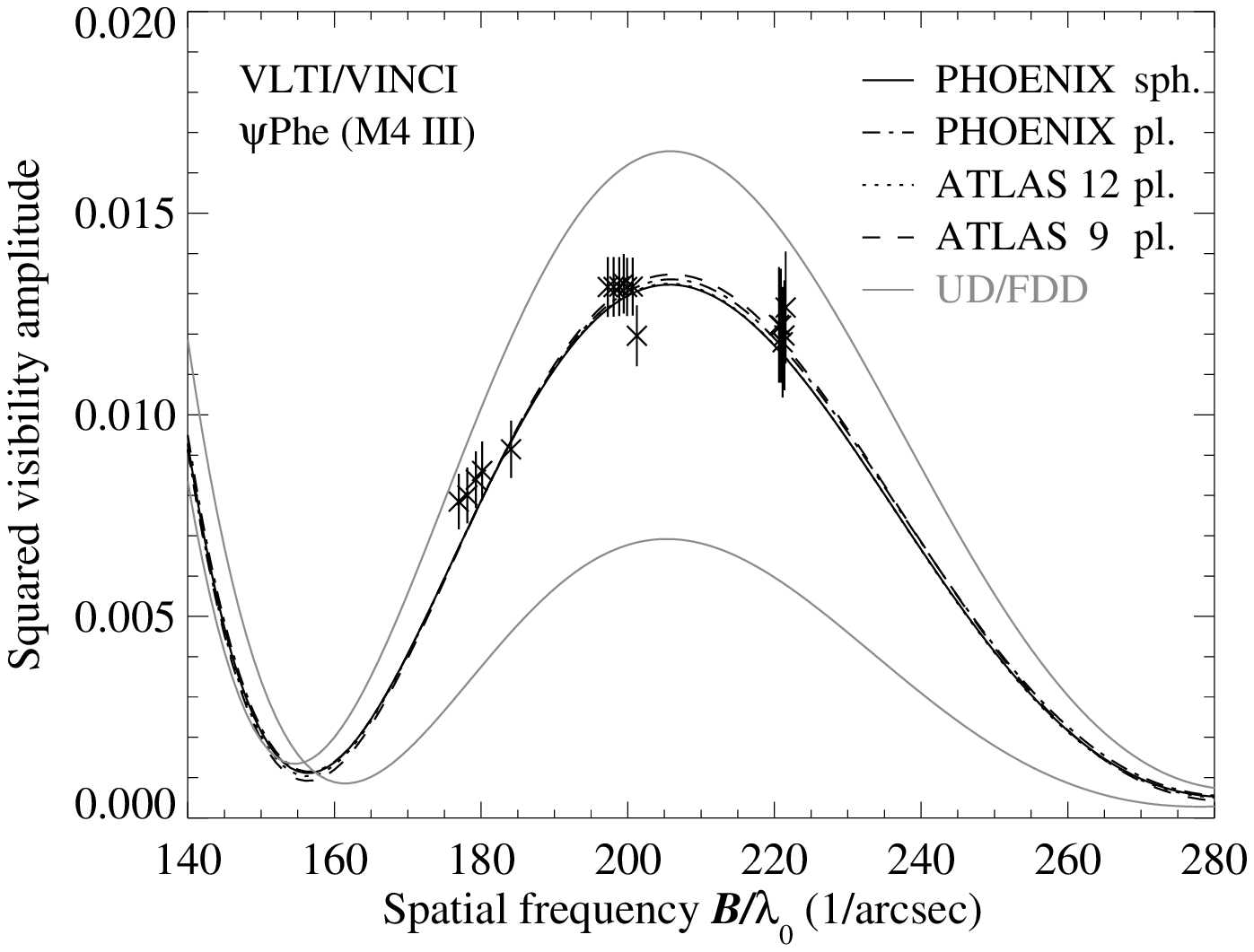}}
\caption{Our measured squared visibility amplitudes of \pp\
('$x$' symbols with error bars) together with the (solid black line) 
spherical {\tt PHOENIX} model prediction with model parameters
\teff, $\log g$, and mass as derived from spectrophotometry
and model evolutionary tracks
(Sect.~\ref{sec:phoenixspec}), and best fitting \adld\ value.
Shown are also the
(dashed-dotted line) plane-parallel {\tt PHOENIX} model,
(dotted line) plane-parallel {\tt ATLAS\,12} model, (dashed line)
plane-parallel {\tt ATLAS\,9} model, all with corresponding model
parameters and best fitting \adld. 
As a reference for the strength of the limb-darkening, the 
gray lines denote corresponding UD (upper line) and
FDD (lower line) model visibility functions.
The left panel shows the full range of the visibility function
while the right panel is an enlargement of the low squared
visibility amplitudes in the second lobe. All considered {\tt PHOENIX}
and {\tt ATLAS} model predictions result in a very similar shape 
of the visibility function in the 2nd lobe.
Our measurements are significantly different from uniform disk 
and fully-darkened disk models, and consistent
with all considered {\tt PHOENIX} and {\tt ATLAS} models.}
\label{fig:vis}
\end{figure*}
All considered {\tt PHOENIX} and {\tt ATLAS} model CLV predictions 
lead to very similar model visibility functions up to the 2nd lobe, 
which are all consistent with our data. Hence, our data confirm the
model-predicted strength of the limb-darkening effect.
Our measured values in the second lobe of the visibility function seem 
to lie systematically above the model predictions by $\sim$\,0.5-1$\sigma$.
It is not yet clear if these small differences are caused by 
systematic effects of our data calibration or by the model structures.
The squared visibility amplitudes 
derived from the siderostat data show a systematic offset of about
1\,$\sigma$ towards larger values with respect to our best fitting curves. 
This is likely caused by possible small systematic calibration effects, 
since the calibration of our high squared visibility amplitudes in the 
range $\sim$\,0.8-0.95 derived from the siderostat data was 
difficult (see Sect.~\ref{sec:calibration}).

The best fitting Rosseland angular diameters are well constrained
by our measurement. The measurements in the 2nd lobe of the visibility
function also constrain the positions of the 1st minimum and 2nd maximum
of the visibility function, which is a constraint of the diameter.
This constraint is independent of possible small systematic
calibration uncertainties of the $|V|^2$ values. 
The reduced $\chi^2_\nu$ values for the different 
considered models do not show significant differences, which is 
consistent with the very similar predicted shapes of the 
model visibility functions. 
These similarities are expected since our measurement is dominated by continuum photons ,
and the continuum forming region of the atmosphere is almost compact
(see Fig.~\ref{fig:udld}).
\section{Discussion and conclusions}
\label{sec:discussion}
\paragraph{Spherical PHOENIX models}
We have constructed a spherical hydrostatic {\tt PHOENIX} model atmosphere
for \pp\ in Sect.~\ref{sec:phoenixspec}. Here, we confront this model's 
prediction for the CLV by comparing it with our VLTI/VINCI measurement
of the visibility function in the second lobe.
We find that the model predicted shape of the visibility function
is consistent with our VLTI/VINCI measurements. 
Simultaneously, the Rosseland angular diameter
derived from the model and spectrophotometry 
$\theta_{\rm Ross}^{\rm Spectr.}=8.0\pm 0.4$\,mas agrees well with
the Rosseland angular diameter derived from the same model and
our VLTI/VINCI measurements 
$\theta_{\rm Ross}^{\rm VINCI}=8.13\pm 0.2$\,mas.
These findings increase confidence in theoretical atmosphere 
modeling of cool giant (M III) stars.

Spherical {\tt PHOENIX} models with varied model parameters
(\teff, $\log g$, $M$) that are still consistent with the values
derived from spectrophotometry in Sect.~\ref{sec:phoenixspec}
lead to the same best fitting values for \adross\ and $\chi^2_\nu$.
Hence, these model parameters cannot be further constrained 
by the shape of our measured visibility function up to the 2nd lobe
beyond the constraints provided by the available spectrophotometry.

The corresponding \adld\ and \adross\ values for these different
spherical models illustrate that \adld\ depends 
on the detailed model structure and that \adross\ is better suited 
to characterize the stellar diameter. The best fitting \adld\ values 
for the different 
considered spherical {\tt PHOENIX} models differ by up to 0.24\,mas or 
$\sim$\,3\%, while the \adross\ values for the same
models differ by only up to 0.02\,mas or $\sim$\,0.2\%.
\paragraph{Plane-parallel {\tt PHOENIX} and {\tt ATLAS} models}
Our plane-parallel {\tt PHOENIX} and {\tt ATLAS} models with the 
same model parameters as used for the spherical model lead, 
as expected for a continuum-dominated measurement, to a 
very similar shape of the visibility function. These small differences 
can not be detected by our VLTI/VINCI measurements. 

The obtained Rosseland angular diameters derived by the plane-parallel
models, which we assume to equal \adld, are
consistent within our error-bars with the result obtained by our 
favorite spherical model.
However, there are systematic differences of the obtained diameter 
values of up to $\sim$\,0.1\,mas or $\sim$\,1.5\% as compared to the 
result obtained by our favourite spherical model. These differences can be
explained by the different model geometries, line lists, 
opacity sampling techniques, and spectral resolutions of the 
employed models.
\paragraph{Final parameter values for \pp}
Because of the 6\% extension of \pp's atmosphere, we consider our
favorite spherical {\tt PHOENIX} model the most reliable one. However,
this can currently not be verified by our VLTI/VINCI measurements.
The diameter \adross\ $=8.13\pm 0.2$\,mas derived from this model 
and our interferometric data is the tightest available constraint 
on \pp's diameter. From this angular diameter and our bolometric flux 
from Sect.~\ref{sec:fbol}, an effective temperature 
of \teff$=3472\pm 125$\,K is derived. This means that the tightest
constraint on \pp's effective temperature comes from the 
model comparison with the available 
spectrophotometry in Sect.~\ref{sec:phoenixspec}, which is 
\teff$=3550\pm 50$\,K. With these values (\adross\ $=8.13\pm 0.2$\,mas, 
\teff$=3550\pm 50$\,K) and the Hipparcos parallax, we derive
a linear Rosseland radius of $R=86\pm 3$\,\rsun\ and a luminosity
of $\log L$/\lsun=3.02 $\pm$ 0.06. Together with the evolutionary tracks
by Girardi et al. (\cite{girardi}), see Sect.~\ref{sec:phot} \&
Fig.~\ref{fig:hr}, these values are consistent with
a mass of $M=1.3\pm 0.2$\,\msun\ and a surface gravity of 
$\log g=0.68^{+0.10}_{-0.11}$. These values are summarized in the last
Col. of Table~\ref{tab:psipheprop}. They are consistent with the
values derived by the different methods mentioned in earlier Sects. 
It would be an interesting further
test to confirm the derived surface gravity and mass by means of 
a high-resolution spectrum. 
\paragraph{Future measurements}
Our VLTI/VINCI measurements could only probe the LD
intensity profile of \pp\ averaged over the broad VINCI sensitivity
band, and confirm the model-predicted strength of the limb-darkening. 
We have shown in Fig.~\ref{fig:udld} that the intensity profile and
the strength of the limb-darkening effect of \pp\ are expected to vary 
over our instrumental bandpass, especially in narrow molecular
bands. While we take this predicted variation into account for the 
computation of the
synthetic broad-band squared visibility amplitudes, 
it would be a better test of the model atmospheres to obtain spectrally 
resolved observations. Promising observations to better constrain 
atmospheres of cool giants would likely be direct limb-darkening
measurements, i.e. observations with more than one resolution element 
across the stellar disk as used here, but at a number of well-defined
narrow molecular and continuum bands with high spectral resolution. 
Theoretical studies seem to be needed and are planned to derive 
which observations in terms of wavelength bands and fundamental parameters 
of the target stars are best suited to constrain model parameters such as
the model geometry and the treatment of molecular opacities. 
The required spectrally resolved limb-darkening observations can
in the future be obtained with the upcoming 
scientific VLTI instruments AMBER and MIDI.
AMBER will allow us to probe the stellar intensity profile with 
a spatial resolution $\lambda/B$ of up to $\sim$\,1\,mas and with 
a spectral resolution $\lambda/\Delta\lambda$ of up to 10\,000. 
AMBER can combine the light from three telescopes simultaneously, and hence 
obtain closure phases and triple amplitudes. The use of closure phases
will likely enable the detection of additional surface features 
such as spots (Wittkowski et al. \cite{wittkowski2}). The use of the
1.8\,m diameter ATs and the 8.2\,m diameter UTs equipped with adaptive
optics will allow us to obtain these measurements with a
high signal-to-noise ratio.
\section{Summary}
We have obtained VLTI/VINCI limb-darkening measurements of \pp\ by
probing the visibility function in the 2nd lobe. Our data confirm the
strength of the limb-darkening effect as well as the Rosseland angular
diameter as predicted by our favorite spherical {\tt PHOENIX} model 
atmosphere, the parameters for which were constrained by
comparison to available spectrophotometry and theoretical 
stellar evolution tracks.
This increases confidence in stellar atmosphere modeling of cool giant 
(M III) stars.

We have derived high-precision fundamental parameters
for \pp, as listed in the last Col. of Table~\ref{tab:psipheprop}.

The measurements presented here are also a precursor for 
planned more detailed spectrally resolved limb-darkening measurements 
of a wider range of stars aiming at further constraining the 
effects of model geometry and the treatment of atomic and molecular lines.
Observations with the completed VLTI and its scientific instruments
will enable these advanced studies of stellar atmospheres.  
\begin{acknowledgements}
Observations with the VLTI are only made possible through the efforts
of the whole VLTI team. We thank especially the VLTI commissioning team 
including the Telescope Instrument Operators 
for the operation of the interferometer. We would like to thank the 
entire PHOENIX model atmosphere team, in particular 
Peter Hauschildt and France Allard for helpful discussions. 
We thank Robert L. Kurucz for making his model atmosphere data
publicly available, for providing us with a new {\tt ATLAS\,12} model
for our source \pp, and for helpful discussions. We thank T. Szeifert
for valuable comments and discussions on this project.
We are grateful for the valuable comments by the referee M. Scholz
which helped to improve this article.
This research has made use of the SIMBAD database, operated at CDS, 
Strasbourg, France.
\end{acknowledgements}
\end{document}